\documentclass[fleqn,10pt]{wlscirep}
\usepackage[utf8]{inputenc}
\usepackage[T1]{fontenc}
\usepackage{float}
\title{Transport Properties and Finite Size Effects in $\beta$-Ga$_2$O$_3$ Thin Films}

\author[1]{Robin Ahrling}
\author[1]{Johannes Boy}
\author[1]{Martin Handwerg}
\author[1]{Olivio Chiatti}
\author[1]{Rüdiger Mitdank}
\author[2]{Günter Wagner}
\author[2]{Zbigniew Galazka}
\author[2,*]{Saskia F. Fischer}
\affil[1]{Novel Materials Group, Humboldt-Universität zu Berlin, Newtonstraße 15, 12489 Berlin, Germany}
\affil[2]{Leibniz Institute for Crystal Growth, Max-Born-Straße 2, 12489 Berlin, Germany}

\affil[*]{sfischer@physik.hu-berlin.de}

\begin{abstract}
Thin films of the wide band gap semiconductor $\beta$-Ga$_2$O$_3$ have a high potential for applications in transparent electronics and high power devices. However, the role of interfaces remains to be explored. Here, we report on fundamental limits of transport properties in thin films. The conductivities, Hall densities and mobilities in thin homoepitaxially MOVPE grown \mbox{(100)-orientated~~$\beta$-Ga$_2$O$_3$} films were measured as a function of temperature and film thickness. At room temperature, the electron mobilities ((115$\pm$10)~$\mathrm{\frac{cm^2}{Vs}}$) in thicker films (> 150 nm) are comparable to the best of bulk. However, the mobility is strongly reduced by more than two orders of magnitude with decreasing film thickness ((5.5$\pm$0.5)~$\mathrm{\frac{cm^2}{Vs}}$ for a 28~nm thin film). We find that the commonly applied classical Fuchs-Sondheimer model does not explain the contribution of electron scattering at the film surfaces sufficiently. Instead, by applying an electron wave model by Bergmann, a contribution to the mobility suppression due to the large de Broglie wavelength in $\beta$-Ga$_2$O$_3$ is proposed as a limiting quantum mechanical size effect.
\end{abstract}

\begin{document}

\flushbottom
\maketitle
%
%
\thispagestyle{empty}

\section*{Introduction}

{Over the past years, gallium oxide (Ga$_2$O$_3$) has proved to be a promising candidate for a variety of possible applications, such as deep UV-detectors, gas sensors, but especially for high power devices.\cite{{Stepanov},{Fortunato},{Marks},{Suzuki}} Gallium oxide is a transparent semiconductor with a high band gap of 4.7~-~4.9~eV at room temperature \cite{{Trippins},{Lorenz},{Orita},{Pearton}}.  The $\beta$-form is most commonly used, since it can be grown from melt and is the most stable configuration. However, to date fundamental material properties in particular for thin films \cite{{Peelaers},{An}} are under investigation.\\
So far, only $n$-type conductivity has been observed in $\beta$-Ga$_2$O$_3$. The effective mass has been reported to be 0.25 - 0.28 electron masses.\cite{{Kang},{Mohamed},{Janowitz}} Various temperature-dependent measurements of electrical parameters on bulk single-crystals, typically grown by the Czochralski-method, have been done before, with the highest mobilities measured in comparable bulk material (Hall densities of about some $10^{17}$~$\mathrm{cm}^{-3}$) so far being  152~$\mathrm{\frac{cm^2}{Vs}}$. \cite{{Galazka2}} With the progress made in the production of homoepitaxial $\beta$-Ga$_2$O$_3$ thin films of high quality, they have a promising prospect for potential use in devices. However, to date, the temperature and film thickness dependence of the electrical behavior of thin films remains largely unknown. \\
In this work, thin homoepitaxially metal-organic vapour phase epitaxy (MOVPE) grown (100) $\beta$-Ga$_2$O$_3$ films were measured. Van-der-Pauw and Hall measurements were taken on the films in a temperature range from 30~K to 300~K for film thicknesses between 28~nm to 225~nm to determine conductivity, Hall density and mobility. Besides electron-phonon and electron-ionized impurity scattering, additional scattering mechanisms for thin $\beta$-Ga$_2$O$_3$ films are discussed to explore their feasibility for future device applications. Especially, a fundamental limit due to finite size effects\cite{Bergmann}, namely the interaction of electron waves with the film boundaries, is explored.\\

\section*{Results}

\subsection*{Bulk and Bulk-Like Thick Films}

One unintentionally doped bulk-$\beta$-Ga$_2$O$_3$ single crystal and several Si-doped homoepitaxial layers were measured. Representative $V(I)$ and $R(B)$ curves for van-der-Pauw and Hall measurements are shown in figure \ref{fig:Probe}. The measured conductivity, Hall density, as well as the mobility of the bulk compared to a 225 nm thick homoepitaxially grown film sample are shown in figure \ref{fig:bulk-werte}.\\

\begin{figure}[H]
\includegraphics[width=.85\textwidth]{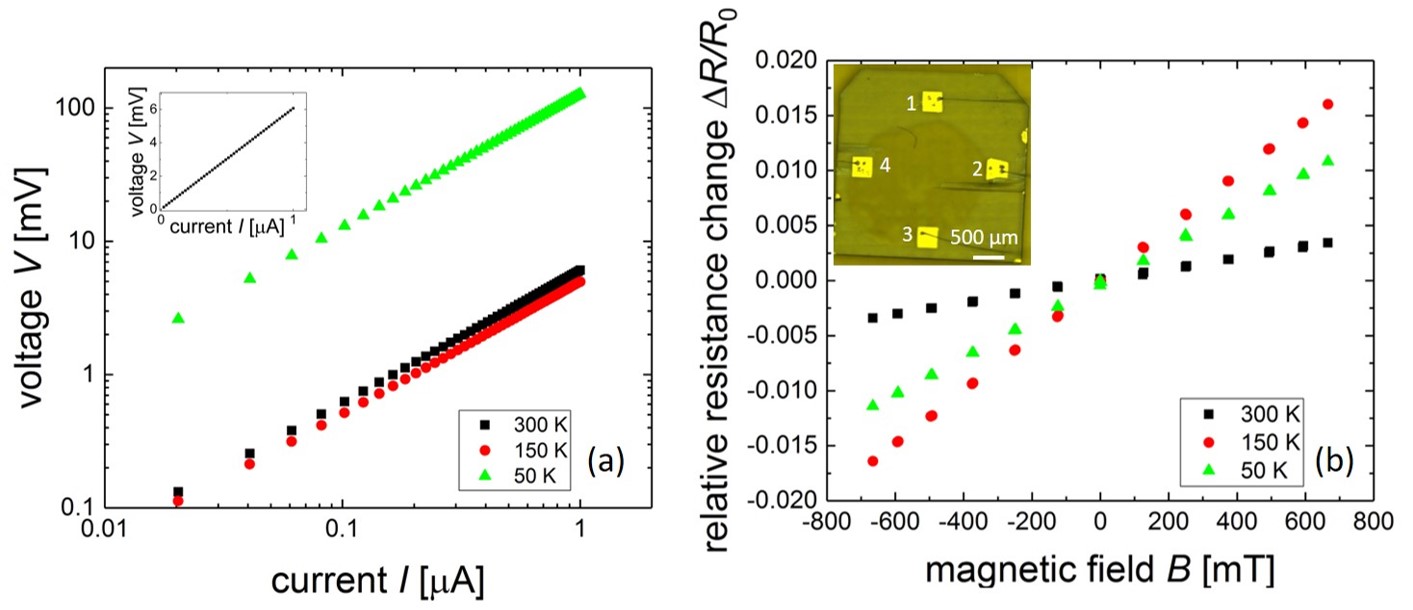}
\caption{\label{fig:Probe} Representative measurement data taken at three different temperatures for \textbf{(a)} van-der-Pauw conductivity measurements, double logarithmic plot \textbf{Inset:} linear plot of 300~K data \textbf{(b)}~Hall measurements. \textbf{Inset:} Representative sample after production of 25~nm~Ti/50~nm~Au contacts and bonding with Al-wire.}
\end{figure}

The conductivity increases with increasing temperature up to a maximum of about 5~S/cm at a temperature of 150~K. For $T \gtrsim 150$~K the conductivity decreases. The Hall density increases strongly with temperature for low temperatures under 150~K. Above that, the increase of the Hall density flattens and saturates for high temperatures. The Hall density in the film is slightly higher than the density in the undoped bulk crystal, which is expected due to the Si-doping.\\
Starting from low temperatures, the mobility shows a strong increase with increasing temperature until it reaches a maximum of about 550~cm$^2$/Vs (350~cm$^2$/Vs) at a temperature of 90~K for the bulk (film) sample. With a further rise in temperature, the mobility decreases. Both samples show an identical temperature dependence, however the film has a slightly lower mobility due to its higher doping level. The mobility values were fit to two dominant scattering mechanisms, ionized impurity scattering for low temperatures and optical phonon scattering for high temperatures.\\
Within the measurement uncertainties the samples with a 4$^\circ$ and 6$^\circ$ off-orientation of the substrate were identical.
The bulk sample and the homoepitaxially grown film are very similar, therefore the same scattering effects are to be expected in both. Since the electrical parameters of the bulk crystal compare well to the best existing literature values\cite{{Irmscher},{Kang},{Oishi}} they can serve as a reference for the following measurements on thin films.

\begin{figure}[ht]

\includegraphics[trim = 0px 0px 3px 0px, clip, width=\textwidth]{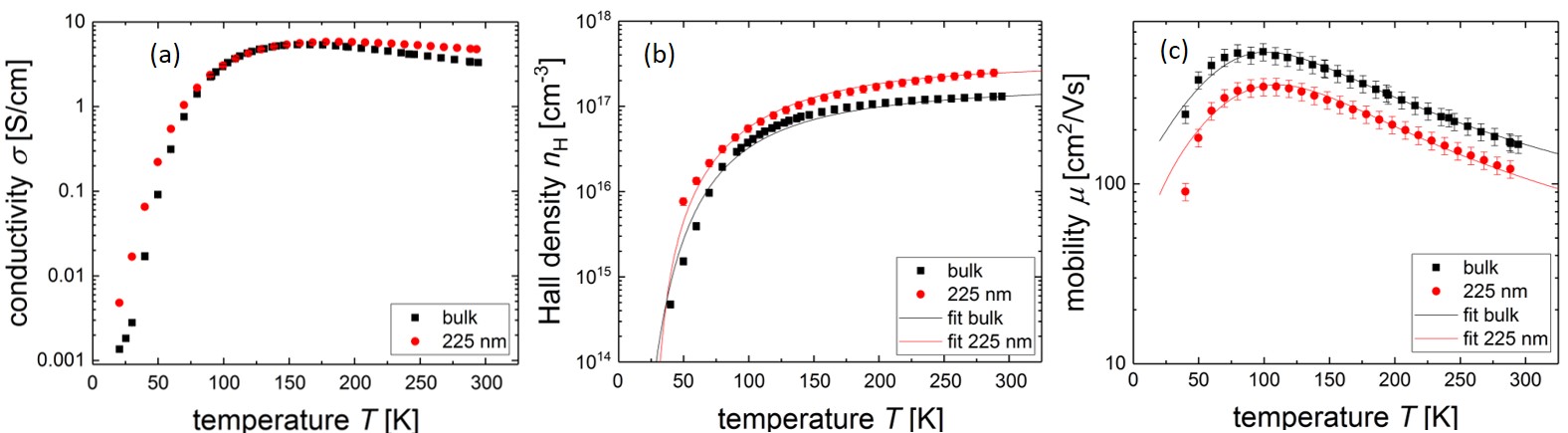}
\caption{\label{fig:bulk-werte} Summary of the reference measurements done on a not intentionally doped, (100)-orientated bulk-$\beta$-Ga$_2$O$_3$ single crystal in direct comparison to a 225 nm homoepitaxially grown Si-doped film. \textbf{(a)} Conductivity $\sigma$ vs. temperature $T$. Both conductivites show a maximum at about 150~K. \textbf{(b)} Hall density $n_\mathrm{H}$ vs. temperature $T$. The data are fit to the charge neutrality equation (see eq. (\ref{eq:charge})) which represents the measured data well. \textbf{(c)} Mobility $\mu$ vs. temperature $T$. The mobilities have been fit to two dominant scattering mechanisms: polar optical phonons, dominant at high temperatures and ionized impurity scattering dominant at low temperatures. Bulk and thin film mobility show a similar form.}

\end{figure}

\subsection*{Thick and Thin Films}

Figure \ref{fig:mu-schichten} shows the measured conductivities, Hall densities, as well as the mobilities of several measured thick and thin film samples. The data of the bulk crystal is also included as a reference.\\
Differences in conductivity and Hall density between the different samples are to be expected due to the doping range ($1\cdot 10^{17}-2\cdot 10^{18}$~cm$^{-3}$). With increasing doping, the donator ionization energy decreases. For higher doping about $2 \cdot 10^{18}$~cm$^{-3}$ the sample approaches the state of a degenerate semiconductor and an almost constant temperature dependence can be observed. Fits were calculated with charge neutrality equation, see eq. (\ref{eq:charge}). As expected, no clear trend in conductivity regarding the film thickness can be seen, however, the differences can be assigned to different doping and sample thicknesses.\\
The thicker films (225~nm - 155~nm) show a mobility function $\mu (T)$ that is very similar to that of the bulk crystal, with a distinct maximum between 100~K and 150~K. This maximum shifts to higher temperatures and lower mobility values if the film thickness is reduced. The absolute mobility values decrease monotonically with decreasing film thickness.\\
The thin films (100~nm and below) show no clear maximum in the $\mu (T)$ function. The curve shape has changed into a monotonic decrease of mobility with decreasing temperature. It should also be noted, that the absolute mobility values for the thin films below 100~nm show a drop up to two orders of magnitude whereas the film thickness only drops by one. This effect can not be ascribed to varying Hall densities. Recently, a similar mobility suppression for thin BaSnO$_3$ films was reported without a detailed analysis.\cite{Sanchela}
In the following, we discuss whether the reason for the observed mobility reduction in thin films is a result of different Hall densities, the crystal quality or the sample thickness.

\subsection*{Temperature Dependence of $n_\mathrm{H}$ and $\mu$}

Regarding the Hall densities $n$, it is shown that the $n(T)$ curves show the expected semiconductor-like temperature dependence, see figure \ref{fig:mu-schichten}b. It is fit with the charge neutrality equation, depending on the acceptor density $N_\mathrm{A}$, the donator density $N_\mathrm{D}$ the donator energy $E_\mathrm{D}$ and the effective mass $m^*$ of the electrons

\begin{eqnarray}
\label{eq:charge}
    \frac{n(n+N_\mathrm{A})}{N_\mathrm{D}-N_\mathrm{A}-n} = N_\mathrm{C}\exp{\left( -\frac{E_\mathrm{D}}{k_\mathrm{B}T} \right)},
\end{eqnarray}

\noindent with the effective density of states $N_\mathrm{C}$ in the conduction band being

\begin{eqnarray}
N_\mathrm{C} = 2 \left( \frac{m^* k_\mathrm{B}T}{2 \pi \hbar} \right)^{3/2} ,
\end{eqnarray}

\noindent $k_\mathrm{B}$ as the Boltzmann constant and $\hbar$ as the reduced Planck constant. Only for the highest doped sample $n(T)$ deviates from the typical semiconductor-like temperature dependence.\\
Comparing the samples with a film thickness of $t=155$~nm and $t=28$~nm shows, that their Hall densities differ only by a factor of 2, whereas the mobility differs by almost 2 orders of magnitude. This suggests an additional scattering mechanism that relates to the sample thickness and becomes dominant in very thin films.\\

Regarding the mobility at high temperatures, no dependence on the film thickness is to be expected for dominating electron phonon scattering. Figure \ref{fig:mu-schichten}c however shows, that there is a clear decrease of mobility with decreasing film thickness. This means, that an additional scattering mechanism must be a function of film thickness.\\
From literature \cite{{Oishi},{Parisini},{Ma},{Ghosh}} it is expected that scattering of electrons with polar optical phonons dominates the high temperature mobility behavior and scattering with ionized impurities dominates the low temperature regime. To test for that, the mobilities are fit to a model considering polar optical phonons and ionized impurity scattering to show the dominant scattering mechanisms. For the fit functions used here, see suppl. inf. S1. Both terms are independent scattering events and can therefore be added to a total mobility $\mu_\mathrm{vol}$ after Matthiessens rule \cite{Matthiessen} if lattice and electron gas are in a thermal equilibrium

\begin{eqnarray}
\label{eq:matt}
\frac{1}{\mu_\mathrm{vol}} = \frac{1}{\mu_\mathrm{OP}}+\frac{1}{\mu_\mathrm{II}}.
\end{eqnarray}

As shown previously for bulk\cite{Oishi} ionized impurity scattering dominates the low temperature behavior of the mobility and optical phonon scattering the high temperature regime. Thicker homoepitaxially grown films above 100~nm thickness show a similar temperature dependence of mobility, indicating that the same scattering mechanisms are dominant. The maximum shifts to higher temperatures and lower mobilities when reducing the film thickness.\\

\begin{figure}[H]
\includegraphics[trim = 0px 0px 0px 5px, clip,width=\textwidth]{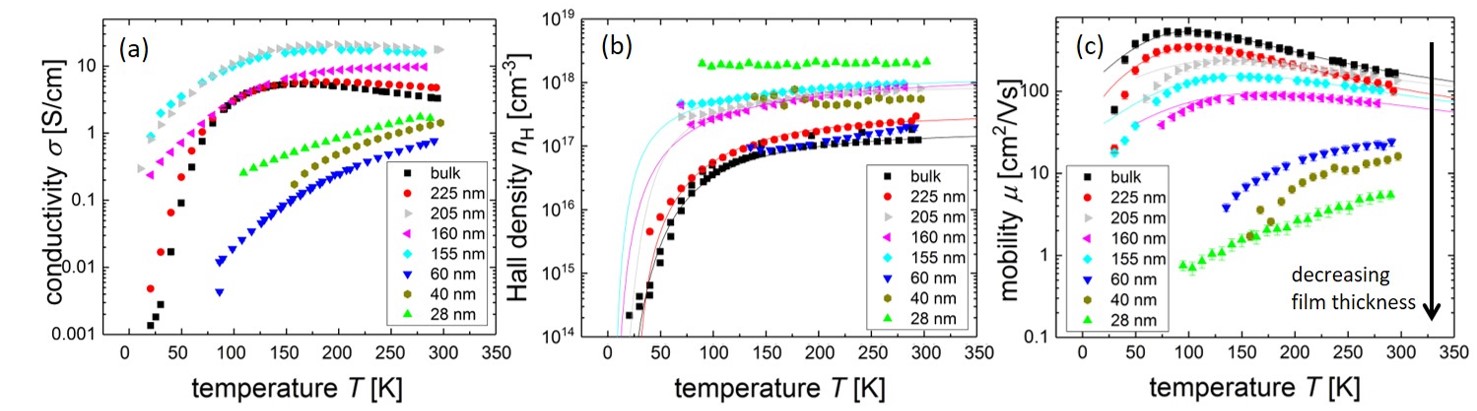}
\caption{\label{fig:mu-schichten} Summary of the measurements done on homoepitaxially grown, (100)-orientated bulk-$\beta$-Ga$_2$O$_3$ Si-doped films (doping range: $1\cdot 10^{17}-2\cdot 10^{18}$~cm$^{-3}$). \textbf{(a)} Conductivity vs. temperature. No clear dependence on film thickness can be observed. \textbf{(b)} Hall density vs. temperature. Fits were carried out with the charge-neutrality equation. \textbf{(c)} Mobility vs. temperature. Thick films (>150~nm) show a behavior similar to the bulk, thin films (<100~nm) show a highly reduced mobility with a different temperature dependence. Note: the Hall density of the 28~nm sample was measured with the AC-Hall method, due to its low mobility.}
\end{figure}

Generally, different doping concentrations lead to different effects on the low temperature mobility due to ionized impurity scattering causing different slopes in the mobility curves for low temperatures. Possible reasons for the deviations of mobility fits from the data points for low temperatures may be neutral impurity scattering in general,  growth problems at substrate/thin film interface (causing a high number of neutral impurities at the interface) or hopping transport due to impurity bands (only expected for densities above $2 \cdot 10^{18}$~cm$^{-3}$). Hopping has been reported to contribute to the conduction in highly doped ($4 \cdot 10^{18}$~cm$^{-3}$) $\beta$-Ga$_2$O$_3$ crystals for temperatures below 150~K \cite{Oishi-II}. For higher temperatures and lower doping such a contribution is not to be expected. For thinner films below 100~nm thickness, a temperature dependence of the mobility with no maximum is observed. There is a drastic reduction in mobility up to two orders of magnitude over the entire measured temperature range. This effect gets stronger with an increased reduction of film thickness. Conducting the same fits on these samples was not successful, showing that different scattering mechanisms play a role here. None of the effects mentioned above is expected to show any direct dependence on the film thickness.

\subsection*{Influence of Twin Boundaries}

Imperfect growth may play a role for the mobility of real samples. The main type of defects in the films examined here are twin boundaries. For the purpose of calculating the scattering of electrons on those defects, the twin boundaries can be treated like grain boundaries in a polycrystalline or powdered semiconductor, with a certain potential barrier that has to be overcome \cite{Seto}. Such a behavior has already been observed for homoepitaxially grown $\beta$-Ga$_2$O$_3$-films \cite{Fiedler}.\\

The scattering of electrons at the twin boundaries adds another term $\mu_\mathrm{tb}$ to the total mobility. This term is temperature dependent and can be described as

\begin{eqnarray}
\mu_\mathrm{tb} = \frac{eL}{\sqrt{8 k_\mathrm{B}T \pi m^\ast}} \exp{\left( - \frac{E_\mathrm{B}}{k_\mathrm{B} T} \right)},
\label{eq:twin}
\end{eqnarray}

where $L$ is the mean distance between twin boundaries and $E_\mathrm{B}$ is the mean potential barrier height.

This model was applied to three representative thin films (28~nm, 60~nm and 155~nm) with $L$ and $E_\mathrm{B}$ as fit parameters. For the 155~nm sample twin boundary scattering was added to the already described phonon and ionized impurity scattering terms, leading to corrections to the mobility fit, most clearly seen at low temperatures, see figure \ref{fig:twin60}. The two thinner films were fit only with the twin boundary model, due to the fact that phonon and ionized impurity scattering do not essentially contribute to the combined scattering. The barrier height is found to be around $E_\mathrm{B}=45$~meV for the 28~nm and 60~nm sample and about $E_\mathrm{B}=20$~meV for the 155 nm sample. $L$ is calculated to be 3~nm, 15~nm and 700~nm for the 28~nm, 60~nm and 155~nm sample respectively.\\
For the two thinner films, the fit represents the data well, even though there are deviations for temperatures below 150~K for the 28~nm sample. However, the strong dependence of the fit parameter $L$ on the film thickness seems unreasonable. Since the crystals defects density is not film thickness-dependent\cite{Schewski}, a systematic reduction of the mobility due to twin boundary scattering with the film thickness is not expected. Grain boundary scattering explains a mobility reduction of films with respect to single crystalline bulk especially for low temperatures. However, it is not sufficient to explain the reduction of two orders of magnitude for thin films. 

\subsection*{Influence of Finite Size Effects}

Assuming films of high quality, possibilities for a mobility reduction in crystallographically perfect thin films are discussed in the following. Characteristic lengths that could play a role in the mobility in thin films are the effective mean free path $l$, the de Broglie wavelength $\lambda_\mathrm{e}$, the sample thickness $t$ and the surface roughness $r_\mathrm{S}$.\\
It has been shown, that depletion due to band bending at the surface does not play a role for free standing thin $\beta$-Ga$_2$O$_3$ films \cite{Peelaers}. However, a possibility for a mobility reduction in thin films is a surface scattering mechanism. In the semi-classical Fuchs-Sondheimer model\cite{{Fuchs},{Sondheimer_alt},{Sondheimer}} for ideal thin metal films the ratio of mean free path to film thickness determines the mobility. Due to the very short electron wavelengths, the situation in metals is typically $\lambda_\mathrm{e} \ll l$ and $l\approx t$, meaning the Fuchs-Sondheimer model describes the scattering well. Furthermore, the electron wavelength is much smaller than the surface roughness $\lambda_\mathrm{e} \ll r_\mathrm{S}$. This leads to diffuse scattering at the surfaces. If these conditions are met, the Fuchs-Sondheimer model can also be successfully applied to thin semiconducting films \cite{Anderson}.\\
However, in the semiconducting films examined here, the conditions are different: $\lambda_\mathrm{e} > l$ and $\lambda_\mathrm{e} \approx t$. An upper limit for $l$ is 6~nm (bulk) at RT. Therefore, it is much smaller than the film thickness even for the thinnest samples and highest temperatures. The scattering described by the Fuchs-Sondheimer model therefore does not play the dominant role as shown in suppl. inf. S2.\\
Instead, the de Broglie wavelength $\lambda_\mathrm{e}$ of electrons in non-degenerate semiconductor films can be approximated by

\begin{eqnarray}
\lambda_\mathrm{e} = \frac{h}{\sqrt{2m^\ast k_\mathrm{B}T}} ,
\end{eqnarray}

leading with $m^\ast =$~$0.28~m_\mathrm{0}, T = 300$~K to $\lambda_\mathrm{e} = 14$~nm. It increases for lower temperatures to tens of nm, therefore finite size effects may play a role when $\lambda_\mathrm{e} \approx t$.\\
A quantum theoretical model by Bergmann \cite{Bergmann} describes the reduction of conductivity, and hence mobility, in a crystallographically perfect thin film due to the interaction of the electron wave with the sample surface (see also  suppl. inf. S3). As an additional scattering mechanism it adds to the total mobility the following term

\begin{eqnarray}
\label{eq:Bergmann}
\mu_\mathrm{Bergmann} = \frac{e}{\hbar} \left( \frac{t}{\lambda_\mathrm{e}} \right)^2 \ln{\left( \frac{t}{\lambda_\mathrm{e}} \right)} \frac{1}{nt}.
\label{eq:dB}
\end{eqnarray}
It mainly depends on the ratio of film thickness $t$ and de Broglie wavelength $\lambda_{e}$ and has no longer any direct dependence on material parameters. \\
All mobility data taken at a temperature of 265~K were plotted against $\mu_\mathrm{Bergmann}$ described in eq. (\ref{eq:Bergmann}), see figure \ref{fig:bergmannfit}a. A fit with 

\begin{eqnarray}
\label{eq:fitgleichung}
\mu_\mathrm{tot} = \left(  \frac{1}{A \mu_\mathrm{Bergmann}}+\frac{1}{\mu_\mathrm{vol}}   \right)^{-1}
\end{eqnarray}

was carried out. Here, $\mu_\mathrm{vol}$ represents all the scattering processes that also take place in the bulk and thick film samples (see eq. (\ref{eq:matt})) and $A$ is a parameter, describing the deviation from the Bergmann model as described in~eq.~(\ref{eq:Bergmann}) .\\

\begin{figure}[H]
\includegraphics[ width=\textwidth]{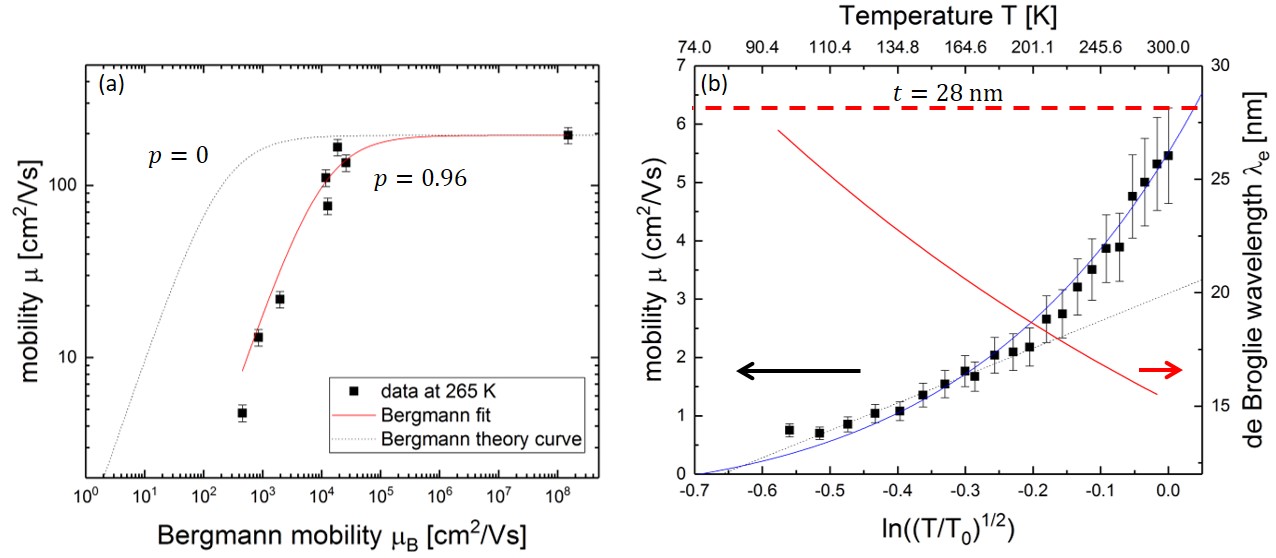}
\caption{\label{fig:bergmannfit} \textbf{(a)} Mobilities of all measured samples vs. the Bergmann mobility. The dotted black line shows the curve directly predicted by the Bergmann model, the solid red line shows a fit with a free parameter $A$, see eq. (\ref{eq:fitgleichung}). The fit represents the data points well, it only deviates by a constant factor of $A=0.02$. \textbf{(b)} Mobility in the 28 nm sample (left axis, black squares)  and de Broglie wavelength (right axis, red line) vs. $\ln{((T/T_0)^{1/2})}$. The solid blue fit curve was done with eq. (\ref{eq:28nm}) and shows good agreement with the data. For temperatures below 150~K, a linear fit was carried out, to show the dominance of the logarithmic $t/\lambda_\mathrm{e,0}$ term. The sample thickness is marked with a thick dashed red line}
\end{figure}

The fit curve in fig.~\ref{fig:bergmannfit}a represents the data points well. However, it deviates from the theoretically predicted curve by a constant factor shifting it to higher values. The deviation can be described by the parameter $A=0.02$.\\
One explanation for this shift is the specularity parameter $p$. The Bergmann model was calculated from the Sondheimer model, but sets $p=0$ since it was originally developed for metals where diffuse surface scattering dominates. In eq. (\ref{eq:Bergmann}) the sheet density $nt$ would scale to $\frac{1+p}{1-p} nt$. For $A=0.02$ the specularity parameter must be $p=0.96$, meaning almost completely specular surface scattering. This can be explained by comparing the de Broglie wavelength in the thin films with their surface roughness. Contrary to the typical situation in metals, here it is $r_\mathrm{S} \ll \lambda_\mathrm{e}$. This was also confirmed by AFM measurements, where the surface roughness was determined to be below 3~nm while the de Broglie wavelength is 14~nm at 300~K. A specularity parameter close to $p=1$ indicating specular scattering is therefore reasonable.\\

Considering this model\cite{Bergmann}, the strongest effect is expected for the thinnest film, the 28 nm sample. The model predicts a temperature dependence of

\begin{eqnarray}
 \mu_\mathrm{Bergmann} &= \mu_0 \frac{T}{T_0} \cdot \ln{\left(\frac{t}{\lambda_\mathrm{e}}\right)} = \mu_0 \frac{T}{T_0} \cdot \ln{\left(\frac{t}{\lambda_\mathrm{e}} \frac{\lambda_\mathrm{e,0}}{\lambda_\mathrm{e,0}} \right)} = \mu_0 \frac{T}{T_0} \left[  \ln{\left(\frac{t}{\lambda_\mathrm{e,0}} \right)} + \ln{\left( \frac{\sqrt{T}}{\sqrt{T_0}} \right)}   \right] .
 \label{eq:28nm}
\end{eqnarray}

Here, $T_0=$~300~K and $\lambda_\mathrm{e,0}= 14$~nm is the corresponding de Broglie wavelength. This temperature dependence is experimentally verified for the 28 nm sample, see figure \ref{fig:bergmannfit}b. It is found, that $\mu_\mathrm{tot}=\mu_\mathrm{Bergmann}$. The data fit to eq. (\ref{eq:28nm}) with a value of \mbox{$\mu_0 = 8$~$\mathrm{\frac{cm^2}{Vs}}$} and shows a good agreement with the fit curve. For low temperatures (150~K and below) the data were approximated linearly to show the dominance of the logarithmic $\ln{(t/\lambda_\mathrm{e,0})}$ term. Extrapolating the fit curve, the mobility reaches zero at about 70~K. Here, the de Broglie wavelength following eq. (\ref{eq:dB}) is \mbox{ $\lambda_\mathrm{e} = 28$~nm}. This is exactly the condition ($t \approx \lambda_\mathrm{e}$) where the mobility should drop to 0 according to the Bergmann model. Generally, formula (\ref{eq:28nm}) approximates the measured data well.\\


\section*{Discussion}

The previous examinations clearly show, that multiple scattering mechanisms play an important role to describe the mobility in the measured homoepitaxial thin $\beta$-Ga$_2$O$_3$ films. We identify three different thickness regimes with respect to dominance of the several scattering mechanisms: The thicker, bulk-like films ($150$~nm and above), an intermediate thickness range ($150 - 50$~nm) and the very thin films ($50$~nm and below).\\
For the bulk-like films, phonon scattering for high temperatures and ionized impurity scattering for low temperatures are the most dominant scattering mechanisms. The scattering at twin boundaries also gives an important correction for low temperatures. A representative fit conducted for the 155~nm sample considering these three mechanisms is shown in figure \ref{fig:twin60}. \\
Both the Bergmann model \cite{Bergmann} and the scattering on twin boundaries \cite{Seto} can mathematically describe the reduction of mobility observed in the thinner films. However, a direct dependence on the film thickness is not to be expected for twin boundary scattering. The mobility reduction due to the Bergmann model clearly dominates in thinnest 28~nm sample, as seen in figure \ref{fig:twin60}. The twin boundary scattering can not explain the low mobilities in thin films alone. A combination of these two mechanisms describes the mobility in the intermediate region between thin and thick films. 

In figure \ref{fig:twin60} the 60 nm sample has been fit to both of these mechanisms and reproduces the measured mobility well. The Bergmann part dominates for higher temperatures and the twin boundary scattering becomes relevant below 150~K. The fit parameters come out to be $\mu_0 = 20 \frac{\mathrm{cm}^2}{\mathrm{Vs}}$, $L=515$~nm and $E_\mathrm{B} = 79$~meV. The de Broglie wavelength $\lambda_\mathrm{e,0} = 14$~nm is independent of the sample thickness. Comparing this to the twin boundary scattering fit results of the 155~nm sample gives a much more consistent development of values for $L$, indicating that the two films are of comparable crystal quality. In the intermediate region between thin and thick films, both mechanisms are relevant and have to be considered. The amount of twin boundaries present in the crystal will determine the exact location of this intermediate region. The different barrier heights $E_B$ can be explained by the different doping levels $N_\mathrm{D} - N_\mathrm{A}$, the potential decreases with with $1/(N_\mathrm{D} - N_\mathrm{A})$ \cite{Seto}. For the samples with a thickness of 155~nm and 60~nm, the data confirm this, as seen in figure \ref{fig:twin60}. For the doping level of the thinnest 28~nm film, the twin boundary scattering plays a negligible role. The Bergmann model reproduces both the film thickness and temperature dependence of the mobility for this sample as can be seen in figures \ref{fig:bergmannfit} and \ref{fig:twin60}. Therefore, it is considered to be the relevant scattering mechanism.\\

\begin{figure}[H]
\includegraphics[trim = 30px 20px 40px 40px, clip,width=.49 \textwidth]{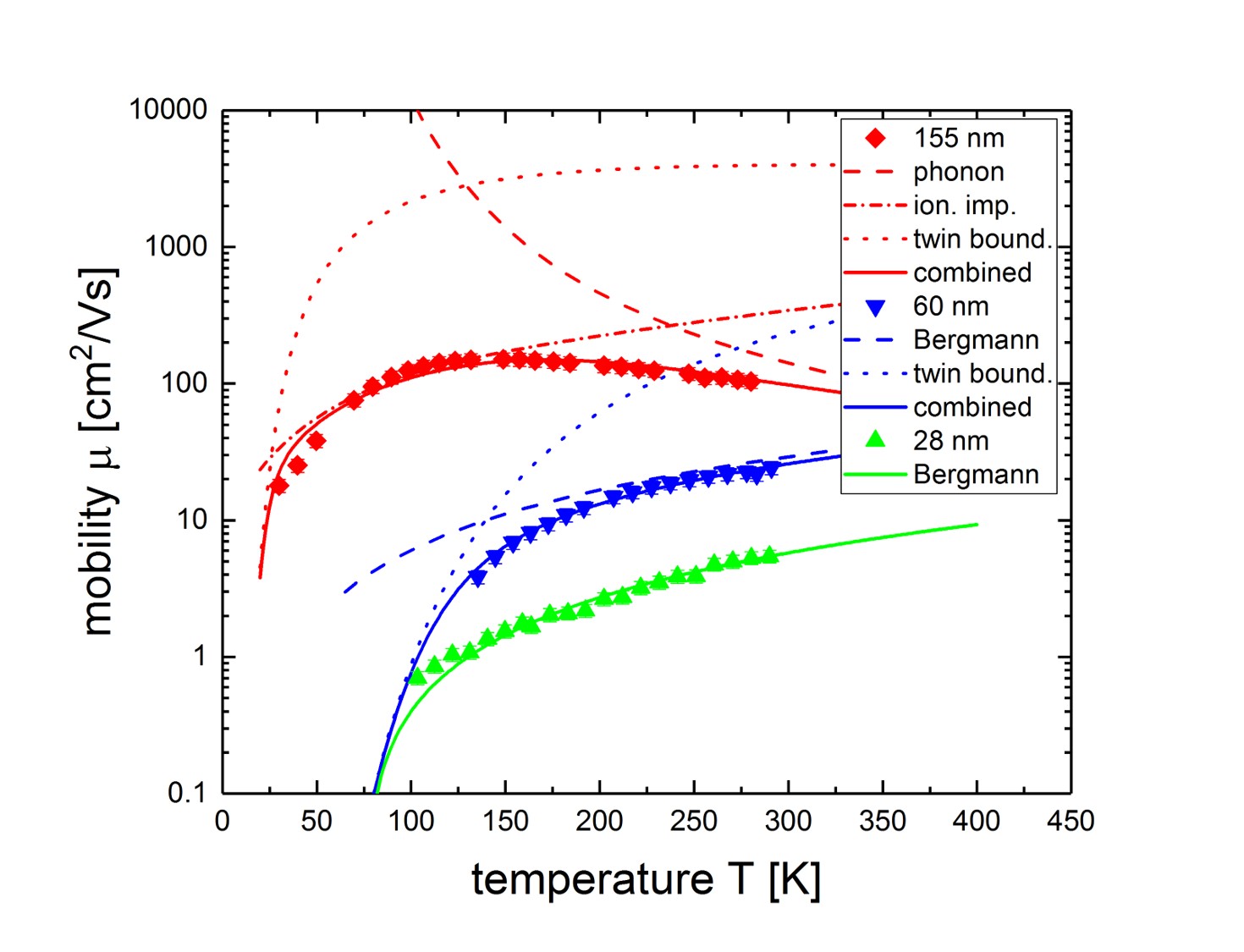}
\caption{\label{fig:twin60} Mobility vs. temperature for selected samples from different thickness ranges. The plot shows an overview of the dominant scattering mechanisms in the three film thickness regimes. Phonon, ionized impurity and twin boundary scattering play a role in the thick films, twin boundary and Bergmann scattering for the intermediate films and for the thin films only the Bergmann part is dominant.}
\end{figure}

In summary, temperature and film thickness dependent measurements on homoepitaxially grown $\beta$-Ga$_2$O$_3$-films were performed in this work. A mobility comparable to bulk crystals was found in the thicker MOVPE-grown films around 200~nm film thickness. Furthermore, the measurements have shown that even the thinnest films (28~nm) have a high enough conductivity to be electrically usable at room temperature. A film thickness-dependent model for the electron mobility was discussed here, considering all dominant scattering mechanisms (optical phonon,  ionized impurity, twin boundary and finite size scattering) for thick bulk-like films, intermediately thick films and thin films. It could be shown, that for the thinnest films, the defect density in the currently grown high quality crystals is not the limiting factor for the mobility. Instead, the mobility here is limited by a finite-size effect of the films. The de Broglie wavelength in these films was found to be comparable to the film thickness. The conductivity can therefore not be described by the typically used Fuchs-Sondheimer model but by the Bergmann model. A quantitative agreement with the data can be seen, when a scaling factor assigned to the specularity parameter is taken into account. For the thicker films however, improvements in crystal growth can still lead to a rise in mobility values.\\

\section*{Methods}

A bulk $\beta$-Ga$_2$O$_3$ single crystal as well as thin homoepitaxial layers have been examined here. Bulk $\beta$-Ga$_2$O$_3$ single crystals were grown from the melt by the Czochralski method \cite{{Galazka},{Galazka2},{Galazka3}} at the Leibniz Institute for Crystal Growth. For that purpose Ir crucibles with an inductive heating were used. For the present study, including substrates for homoepitaxy, 2 inch diameter crystals were obtained, which required a high oxygen concentration in a growth atmosphere (supplied in a specific way) to overcome thermodynamic limitations \cite{Galazka3}. The crystals were either electrically insulating (doped with Mg), or semiconducting (undoped), which all were grown along the [010] crystallographic direction. Mg-doped crystals were used for off-oriented (100) substrate preparation for the growth of homoepitaxial layers. The substrate preparation was done by CrysTec GmbH, Berlin.\\
Thin films were grown via MOVPE\cite{Baldini}. Trimethylgallium and pure oxygen were used as precursors. \mbox{Tetraethylorthosilicate} was used as source fir Si doping. The substrate temperature during deposition was kept of about 825$^\circ$C, the chamber pressure was set to 5~mbar.\\
The electrically insulating (100)-orientated Mg-doped \mbox{$\beta$-Ga$_2$O$_3$} single crystals were used as substrate for the MOVPE growth of the electrically conducting epitaxial Si-doped $\beta$-Ga$_2$O$_3$ films, meaning the epitaxial layers were grown in (100)-orientation as well. The (100) surface of the substrates were prepared with a 4$^\circ$-6$^\circ$ off orientation from the [100] axis in [00-1] direction to reduce island growth in the films. \cite{Wagner}. To enhance their $n$-type semiconductor characteristics, the films were doped by silicon. Typically, the substrates/films had a size of 5~mm~x~5~mm with film thicknesses varying between 28~nm and 225~nm.\\
It has been shown by AFM measurements \cite{Schewski} that the MOCVD grown (100) $\beta$-Ga$_2$O$_3$-films with a substrate off-orientation angle between 4$^\circ$ and 6$^\circ$ exhibit step flow growth. Also, no interface could be observed between substrate and film in TEM pictures\cite{Wagner}, proving homoepitaxial growth. This was also confirmed by diffraction patterns. This gives reason to assume that the grown films generally are of high structural quality. If defects, mainly a low density of twin boundaries, exist, they grow through the whole film \cite{Schewski}, meaning that there is no  systematic difference between samples of different thicknesses to be expected.  However, the interface between substrate and film is strongly defined by their different conductivities. The substrate is insulating whereas the film is n-conducting, leading to a potential difference at the interface.\\
No significant difference in sample surface roughness between the thickest (225~nm) and the thinnest measured film (28~nm) have been observed by AFM measurements. Both surfaces showed an average roughness of well below 3~nm.\\

Ohmic contacts on the sample were realized using aluminum (Al), gold (Au) and titanium (Ti). After cleaning the samples in acetone, a positive photoresist (AZECI 3007) was spun on to the sample at 3500~rpm for 50~s. After that, the laser lithography was performed and the sample was developed in a developer solution (AZ 326 MIF). The metal contacts were then produced via magnetron sputtering of 25~nm Ti and 50~nm of Au onto the sample with subsequent liftoff in acetone in an ultrasonic bath. In a next step, the sample was glued into a chip carrier for later measurements and wedge bonded with Al-wire. It should be noted, that ohmic contacts were achieved only by bonding of those contacts with the Al-wire, therefore point like contacts on the bonding sites can be realized.

All measurements were done in a KONTI IT\texttrademark flow cryostat. Van-der-Pauw and Hall-measurements were carried out in a temperature range between 30~K and 300~K. For the van-der-Pauw measurements, the sample is ideally contacted at four points very close to the sample edges. A current $I$ is then applied between the two neighboring contacts (1,2), the voltage $V$ is measured along the other two contacts (3,4). Doing this measurement in two alternating configurations yields two resistances ($R_{12,43}, R_{23,14}$), from which the conductivity $\sigma$ can be determined by

\begin{eqnarray}
\frac{1}{\sigma} = \frac{\pi t}{\ln{(2)}}\frac{R_{12,43} + R_{23,14}}{2}f,
\label{eq:vdp}
\end{eqnarray}
where $t$ is the film thickness and $f$ is a correction factor depending on the sample shape. The deviation caused by the placement of contacts away from the sample edges leads to another correction factor, needed to calculate the correct conductivity and Hall density. Those correction factors were simulated with a finite elements simulation using the program \mbox{COMSOL\texttrademark}.\\
The measurements were taken with a Keithley\texttrademark 2450 Sourcemeter, evaluating a current sweep to obtain the resistances being measured. With these measurements, conductivities and Hall carrier densities were determined. The Hall coefficient is assumed to be 1 in this work. The Hall mobility $\mu_\mathrm{H}$ is therefore written simply as $\mu$. The Hall density $n_\mathrm{H}$ can be determined, when measuring the sample in a magnetic field $B$ perpendicular to the sample. With variations in the magnetic field $\Delta B$, the measured resistance $R_{13,24}$ will change by $\Delta R$ and the Hall density can be calculated as

\begin{eqnarray}
n_\mathrm{H} = \frac{\Delta B}{et\Delta R},
\end{eqnarray}

with $e$ as electron charge and t as film thickness. With the conductivity $\sigma$ and the Hall density the mobility $\mu$ could be calculated using

\begin{eqnarray}
\label{eq:mu}
\mu = \frac{\sigma}{en_\mathrm{H}}.
\end{eqnarray}

From the mobiities, an effective mean free path $l$ for the carriers can be calculated as

\begin{eqnarray}
l = \frac{\mu}{e}\sqrt{2m^\ast k_\mathrm{B}T},
\end{eqnarray}
with $m^\ast$ as effective mass. For samples with very low mobilities (below 1~cm$^2$/Vs), an AC measurement technique was used to determine the Hall densityby using periodically oscillating magnetic fields (frequency below 1~Hz). A similar approach was used for example by Chen \textit{et al.} \cite{Chen}.

\bibliography{ManuskriptSR}

\section*{Acknowledgements}

This work was performed in the framework of GraFOx, a Leibniz-ScienceCampus, partially funded by the Leibniz association and by the Deutsche Forschungsgemeinschaft (FI932/10 - 1 and FI932/11 - 1). The authors would like to thank Maximilian Kockert and Robert Schewski for fruitful scientific discussions, Raimund Grünberg for technical help in MOVPE growth, and Max Pfeifer for assistance in the experimental work.

\section*{Author contributions statement}

Z.G. conducted the bulk crystal growth, G.W. conducted the MOVPE growth, R.A., J.B., M.H., O.C., R.M. and S.F.F. contributed to transport experiments and analyzed the data, R.A., R.M. and S.F.F. wrote the manuscript and all authors contributed to it.

\section*{Additional information}

\textbf{Competing interests:} The authors declare no competing interests. 

\newpage

\section*{Supplementary Information}

\subsection*{S1 Scattering by Optical Phonons and Ionized Impurities}

The equations to fit the different scattering mechanisms were used as follows (for more details see Oishi \textit{et al.} \cite{Oishi}): The scattering of electrons with optical phonons is described as

\begin{eqnarray}
\label{eq:mu_OP}
\mu_{\mathrm{OP}} &= \frac{4 \varepsilon_0 \pi \hbar^2 \left[ \exp{\frac{E_\mathrm{OP}}{k_\mathrm{B}T}-1} \right]  \left( 1-5 \frac{k_\mathrm{B}T}{E_\mathrm{g}} \right) }{e m^\ast \left( \frac{1}{\varepsilon_\infty}-\frac{1}{\varepsilon_\mathrm{S}} \right) \sqrt{2m^\ast E_\mathrm{OP} \left( 1+\frac{E_\mathrm{OP}}{E_\mathrm{g}} \right)}} ,
\end{eqnarray}
where $E_\mathrm{g}$ describes the gap energy, $E_\mathrm{OP}$ the energy of the optical phonons, $\varepsilon_0$ the vacuum dielectric constant, $\varepsilon_\mathrm{S}$ the low frequency dielectric constant and $\varepsilon_\infty$ the high frequency dielectric constant.\\
The scattering of electrons with ionized impurities is described as

\begin{eqnarray}
\label{eq:mu_II}
\mu_\mathrm{II} &= \frac{128 \sqrt{2} \varepsilon_\mathrm{S}^2 (k_\mathrm{B}T)^{3/2}}{\sqrt{m^\ast}Z^2e^3N_\mathrm{II}} \left( \ln{(1+b)}-\frac{b}{1+b}  \right)^{-1} ,
\end{eqnarray}

with $N_\mathrm{II}$ as density of ionized impurities, $Z$ as the electrical charge of the scattering centers and $b$ being

\begin{eqnarray}
b = \frac{96 \pi^2 \varepsilon_\mathrm{S} \varepsilon_0 m^\ast}{N_\mathrm{II}} \left( \frac{k_\mathrm{B} T}{\hbar e} \right)^2
\end{eqnarray}

\subsection*{S2 Application of the Fuchs-Sondheimer Model}

The commonly used model to explain a reduction in mobility in thin metal films as a function of film thickness is the Fuchs-Sondheimer model. For the case of $l \ll t $ it can be approximated to \cite{{Sondheimer_alt},{Sondheimer}}
\begin{eqnarray}
\label{eq:FS}
\frac{\mu_\mathrm{f}}{\mu_\mathrm{bulk}} = \frac{1}{1+\frac{3l}{8t}(1-p)},
\end{eqnarray}
describing the ratio of the mobility in thin films $\mu_\mathrm{f}$  and in bulk $\mu_\mathrm{bulk}$ in dependence of the ratio of the mean free path $l$ and the film thickness $t$. $p$ denotes the specularity parameter, ranging from $p=0$ for completely diffusive and $p=1$ for completely specular scattering. The maximum mobility reduction takes place when scattering is completely diffusive.\\
As can be seen in figure~\ref{fig:FSfit}, a fit with eq.~\ref{eq:FS} was carried out. Even for the most extreme assumptions, a specularity parameter of $p=0$ and a constant mean free path of bulk of $l_\mathrm{const}=6$~nm, the Fuchs-Sondheimer model can not explain the mobility reduction observed in the homoepitaxial $\beta$-Ga$_2$O$_3$ thin films.

\begin{figure}[ht]
\includegraphics[trim = 30px 20px 40px 40px, clip,width=.49 \textwidth]{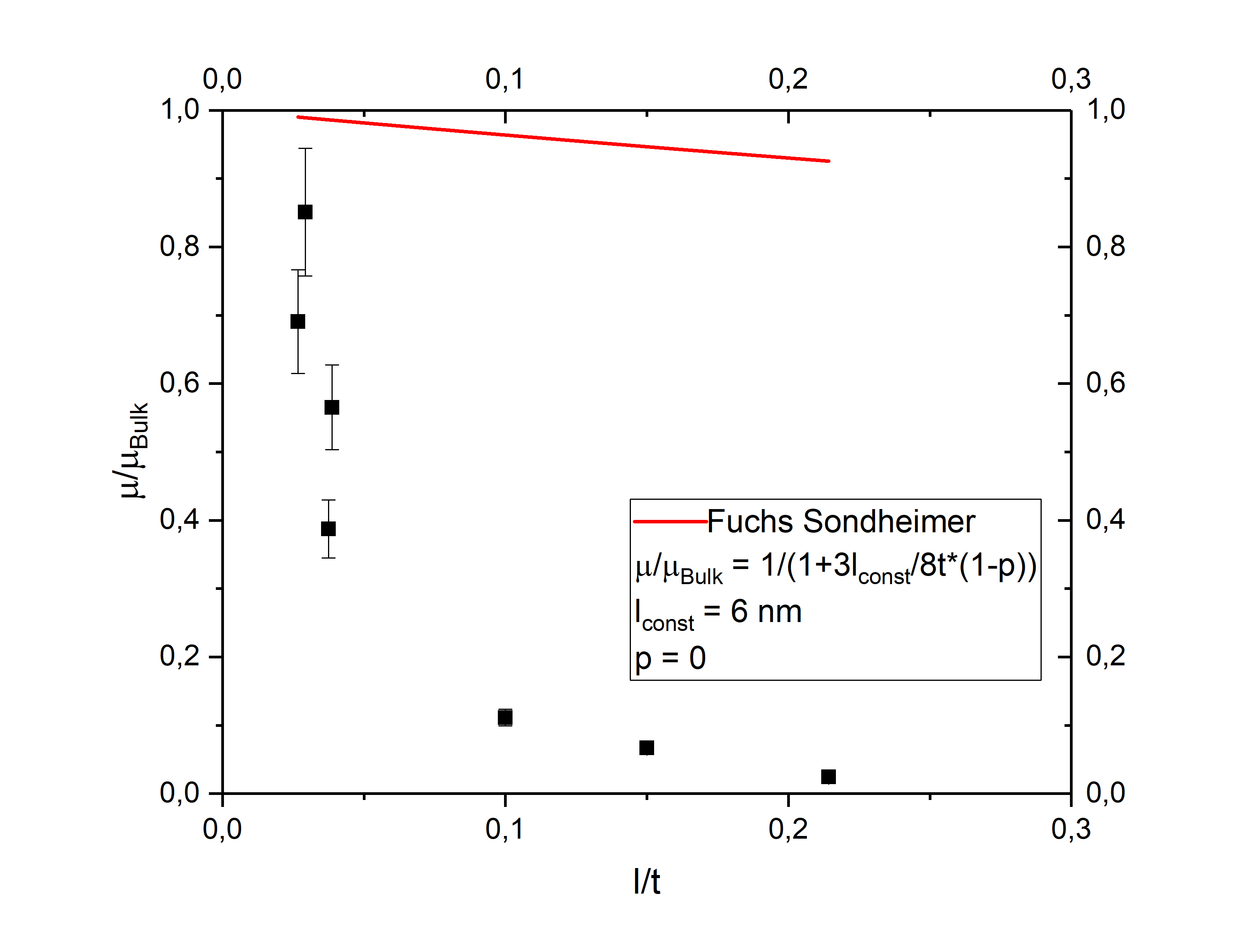}
\caption{\label{fig:FSfit} Thin film mobility relative to bulk mobility vs. ratio of mean free path to film thickness. The red line shows a fit after the Fuchs-Sondheimer model for a constant value of $l=6$~nm. The observed mobility reduction is much larger than the model could explain.}
\end{figure}
  
\subsection*{S3 Application of the Bergmann Model}  
  
In contrast to the Fuchs-Sondheimer model, the Bergmann model \cite{Bergmann} also takes into account the wave properties of electrons. It was developed for samples with an infinite intrinsic mean free path and describes a different dependence for the mobility reduction with decreasing film thickness. Here, the wavelength of the electrons plays an important role. In contrast to the intrinsic mean free path $l_\mathrm{int}$, the effective mean free path $l$ in the Bergmann model is not infinite but is reduced as the mobility gets reduced. Since the electron wavelength becomes comparable to the film thickness in the thin semiconducting samples examined in this work, the wave nature of the electrons becomes important and the Bergmann model can be considered for the explanation of the mobility suppression. 

\end{document}